\documentclass[%
preprint,
 amsmath,amssymb,
 aps,
prb,
 longbibliography,
 lengthcheck,%
]{revtex4-1}

\usepackage{graphicx}
\usepackage{dcolumn}
\usepackage{bm}
\usepackage{color}
\usepackage{hyperref}

\begin{document}


\title{Appearance of a Domain Structure and its Electronic states in Iron Doped 1$T$-TaS$_2$ Observed using Scanning Tunneling Microscopy and Spectroscopy}

\author{Yuita Fujisawa$^{1}$}
\email{1215702@ed.tus.ac.jp}
\author{Tatsunari Shimabukuro$^{1}$, Hiroyuki Kojima$^{1}$, Kai Kobayashi$^{1}$, Shun Ohta$^{1}$, Tadashi Machida$^{2}$, Satoshi Demura$^{1}$}

\author{Hideaki Sakata$^{1}$}

\affiliation{\\$^{1}$Department of Physics, Tokyo University of Science, 1-3 Kagurazaka, Shinjuku-ku, Tokyo 162-8601, Japan \\
$^{2}$Riken Center for Emergent Matter Science, Wako, Saitama 351-0198, Japan}
\date{\today}

\begin{abstract}
In this paper, we report on scanning tunneling microscopy and spectroscopy (STM/STS) measurements on iron doped 1$T$-Ta$_{1-x}$Fe$_x$S$_2$.
A novel domain structure composed of the domains with localized nature divided by the walls is observed 
in Ta$_{0.99}$Fe$_{0.01}$S$_2$, where the Mott transition is completely suppressed.
This indicates that the melting of the Mott state accompanies the appearance of the domain structure.
Since the number of walls increase in superconducting Ta$_{0.98}$Fe$_{0.02}$S$_2$, 
the domain walls seem to be responsible for superconductivity in iron doped 1$T$-TaS$_2$.
\end{abstract}

\pacs{Valid PACS appear here}
\maketitle
Mott physics has attracted much attention because it provides a stage of exotic electronic states such as high temperature superconductivity.
Exotic electronic states usually appear when a Mott state melts into a metallic state.
For example, in cuprates, high temperature superconductivity appears when the Mott state is melted by carrier doping.
Thus, the melting of the Mott state is a profound problem in Mott physics.
The Mott state has been observed in several oxides.
Quite different systems show the Mott state.
One of such systems is a transition metal dichalcogenide, 1$T$-TaS$_2$ (Fig. 1(a)).

1$T$-TaS$_2$ undergoes several charge density wave (CDW) states that are strongly tied to the electronic features, including the Mott state.
Below 180 K, 1$T$-TaS$_2$ shows a commensurate CDW (CCDW) state. 
In the CCDW state, the so called  `David-star' cluster composed of 13 Ta atoms forms a triangular lattice, as shown in Fig. 1(b) schematically.
Simultaneously, the Mott insulating state emerges owing to strong correlation effect of the Ta 5$d$ electrons at the center of each David-star\cite{Hasegawa}.
On warming above about 200 K, the CCDW state transforms into the nearly commensurate CDW (NCCDW) state.
In this state, the spatially uniform CCDW is divided into periodic domains by domain walls called discommensurations\cite{NCCDW_STM}.
Simultaneously, the Mott state melts into a metallic state.
At 600 K, 1$T$-TaS$_2$ undergoes a phase transition from an incommensurate CDW state to a normal state.
Thus, the Mott insulating state in this material is strongly connected with the change in the CDW state.

\begin{figure}[tbh]
\begin{center}
\includegraphics[width=8cm]{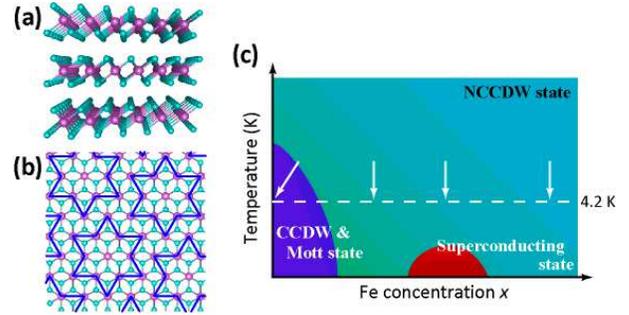}
\end{center}
\caption{(Color online) (a) The crystal structure of 1$T$-TaS$_2$ in the normal state drawn by VESTA\cite{VESTA}.
(b) Scheme of the CDW formation below 350 K. Each cluster illustrated by blue lines is called as the `David-star' cluster.
(c) The Schematic electronic phase diagram of Fe doped 1$T$-TaS$_2$.
Four arrows in the figure represent the Fe concentration in which we have performed STM/STS measurements.}
\label{fig1}
\end{figure}

The melting of the Mott state in 1$T$-TaS$_2$ is not only caused by the increase in temperature.
Sipos {\it{et al.}} reported that application of hydrostatic pressure suppressed the Mott state \cite{Pressure}.
Furthermore, superconductivity was found to emerge up to high hydrostatic pressure. 
Thus, the melting of the Mott state in 1$T$-TaS$_2$ realizes exotic electronic states, such as superconductivity.
They argued the appearance of a textured CDW state, which consists of metallic interdomain spaces and CCDW domains when the Mott state melts.
Substitution of constituent elements also suppresses the Mott state, and induces superconductivity.
Only substitution of a few percent of Fe was reported to reduce the Mott state and cause superconductivity \cite{FeDoping}.
ARPES measurement claimed the real space coexistence of the CDW and superconductivity in Fe substituted 1$T$-TaS$_2$ \cite{ARPES}.
Substitution of anion also causes the melting of the Mott state and the appearance of superconductivity \cite{SeDoping}.
Strong external perturbations also melt the Mott state.
An excitation by a short laser pulse causes metastable ``hidden state," which is inaccessible under normal equilibrium condition \cite{laser1, laser2}.
In this state, spatial ordering of polaron has been argued.
As these reports show, it is thought that the melting of the Mott state in 1$T$-TaS$_2$ accompanies a spatial pattern caused by the change in the CDW state.
Thus, it is interesting to investigate how the spatial pattern evolves with melting in real space.
\begin{figure*}[tbh]
\begin{center}
\includegraphics[width=16cm]{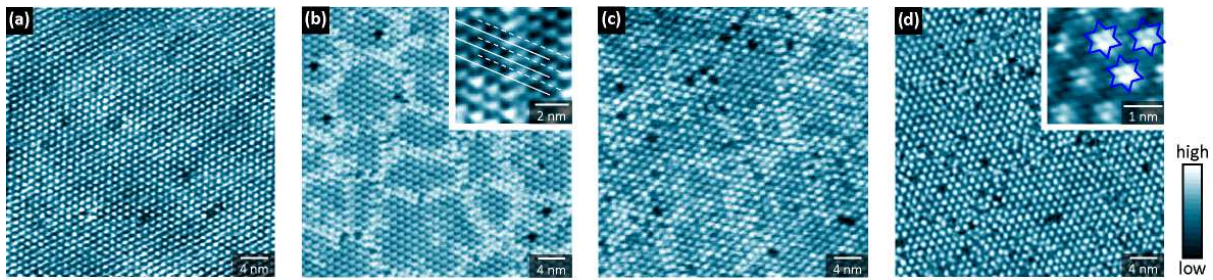}
\end{center}
\caption{(Color online) Typical STM topographs of 1$T$-Ta$_{1-x}$Fe$_x$S$_2$ ($x$ = 0, 0.01, 0.02, and 0.05), respectively.
The set point is 500 mV and 200 pA.
}
\label{fig2}
\end{figure*}

Recently, an application of a short voltage pulse from a tip of a scanning tunneling microscope was reported to realize a metallic metastable state \cite {pulseSTM1, pulseSTM2}.
The metallic metastable state is accompanied by an anomalous domain structure called the ``metallic mosaic" \cite{pulseSTM1, pulseSTM2}.
How this spatial pattern observed in the metastable state relates to the one at the melting of the Mott state by the other techniques such as Fe substitution is an interesting problem.
In this paper, we describe experiments using scanning tunneling microscopy and spectroscopy (STM/STS) on 1$T$-TaS$_2$ with different Fe concentrations to unveil the melting of the Mott state by Fe doping.

Single crystals used in this STM/STS study were grown using a chemical vapor transport method.
We prepared samples with nominal concentrations $x$ = 0, 0.01, 0.02, and 0.05.
Electrical resistivity measurements revealed that the $x$ = 0 sample showed the phase transition from the NCCDW state to the CCDW state at 180 K,
whereas $x$= 0.01 - 0.05 samples did not.
At $x$ = 0.02, the sample showed superconducting transition at 3 K.
The sample with $x$ = 0.05 did not show superconductivity, and insted exhibited an increase in resistivity due to the Anderson localization \cite{AndersonLocalization}.
The phase diagram against Fe concentration is shown in Fig. 1(c).
STM and STS measurements were performed at 4.2 K in the He gas environment with a laboratory-made scanning tunneling microscope.
A clean surface of Ta$_{1-x}$Fe$_x$S$_2$ was obtained by cleavage at 4.2 K.
Electrochemically polished Au wire was used as an STM tip.
STM topographs were taken in the constant current mode.
Tunneling spectra were obtained by the numerical differentiation of the $I$-$V$ characteristics.

The main panels in Fig. 2 represent typical STM topographs of each sample taken at 40 $\times$ 40 nm$^2$ field of view.
As shown in the inset of Fig. 2(d), surface S atoms modulate and form the David-stars, and an array of David-stars covers all the surface in each sample.
These topographs suggest that the David-star itself is stable against Fe doping even at $x$ = 0.05, consistent with a previous STM study \cite{STMTaFeS2}.
In all topographs including  $x$ = 0, several defects can be seen as black points.
Although the number of the defects slightly increases with Fe doping, the number is much lower than Fe concentrations.
Thus, these defects are not thought to be responsible for substituted Fe, but are thought responsible for defects of constituent elements.

The STM topograph at $x$ = 0, where the sample is in the CCDW state, shows the triangle regular array of David-stars. 
With Fe doping, the array of David-stars changes its configuration.
When 1\% Fe is doped, the regular array of David-stars is divided into domains by domain walls.
The size of domains is approximately 100 nm$^2$.
The domain walls show white contrast in the STM topograph at positive bias voltage.
The magnified image is shown in the inset of Fig. 2(b).
The wall is composed of a one dimensional array of two David-stars, which are slightly closer to each other than in the domain.
Because of this structural change, the phase of CDW, {\it{i.e.}}, the phase of the array of David-stars changes across the domain wall.
The NCCDW state in pure 1$T$-TaS$_2$ shows periodic domains of the CCDW divided by discommensurations.
Each discommensuration shows dark contrast compared to the commensurate domains in the STM topograph obtained with positive bias voltage\cite{NCCDW_STM}.
In contrast, the domain structure obtained at $x$ = 0.01 shows irregular domains.
Furthermore, discommensurations show bright contrast compared with domains.
This indicates that the observed domain structure is quite different from the NCCDW state from both the structural and electronic points of view.

At $x$ = 0.02, there exist discrete domain walls and white contrasted David-stars.
Because the sample shows superconductivity,
the topographic change must be related to the emergence of superconductivity.
We will discuss the cause of this change and its relationship with superconductivity later.
At $x$ = 0.05, the array of David-stars is so distorted that the domain walls at which the phase change of the array is well defined no longer exist as shown in Fig. 2(d).
Such a distortion seems to result in logarithmical increase in resistivity behavior \cite{AndersonLocalization}.

\begin{figure*}[tbh]
\begin{center}
\includegraphics[width=16cm]{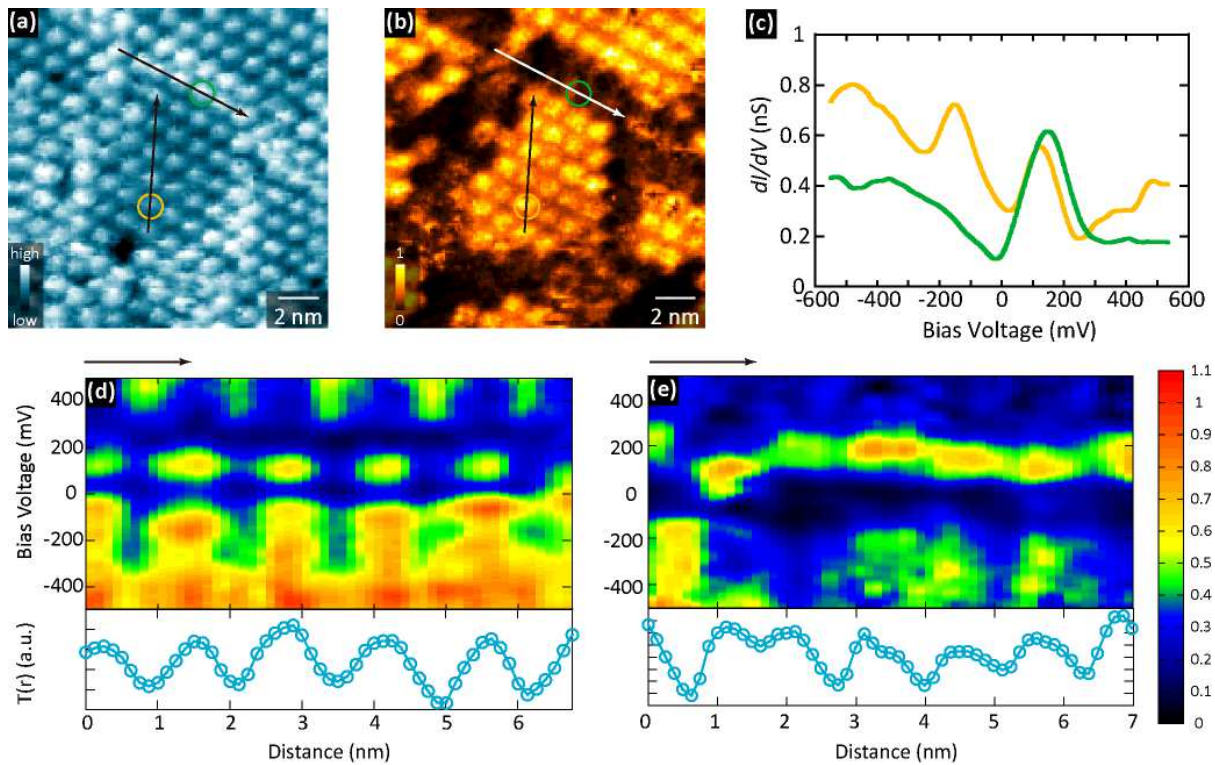}
\end{center}
\caption{
(Color online) STS results on the domain structure. The set point is 500 mV and 200 pA.
(a), (b); An STM topograph and a simultaneously obtained differential conductance ($dI/dV$) map at -100 mV.
(c) Typical tunneling spectra obtained at the David-star in the domain (orange) and the domain wall (green), shown as circles in (a) and (b).
(d), (e); Tunneling spectra along the arrows in the domain and the domain wall, respectively. 
Lower plots represent the line profiles of the STM topograph along the arrows in the domain and the domain wall.
}
\label{fig3}
\end{figure*}

These observations revealed that Fe doping drastically changes the CDW configuration.
Especially, the Mott state was found to melt through the appearance of the domain structure.
Next, we examine the electronic states of the domain structure by STS.
Figures 3(a) and (b) show the STM topograph and the differential conductance ($dI/dV$) map at -100 mV at the same field of view in the sample with $x$ = 0.01.
The domain can be imaged as bright contrast in the $dI/dV$ map, indicating different electronic states between the domain and the wall.
Figure 3(c) shows the tunneling spectrum obtained at the David-stars in the domain (orange).
The tunneling spectrum obtained in the domain shows two peaks at approximately $\pm$100 mV, which seems to be a reminiscent of the Mott gap\cite{Hasegawa}.

Figure 3(d) shows the change in the tunneling spectra along the arrow in the domain shown in Figs. 3(b).
The conductance is shown in color scale.
The tunneling conductance at $\pm$100 mV along the arrow is found to show periodic change.
The lower graph shows the line profile of the STM topograph.
Each maximum indicates the position of the David-stars.
The conductance is found to show maxima at the top of each David-star.
In the Mott insulating state at $x$ = 0, because each electron at the center of each David-star is localized owing to the strong correlation,
the density of states (DOS) derived from the lower and the upper Hubbard band are predominantly enhanced at the center of each David-star\cite{Hasegawa}.
The observed feature in the domain is similar to this situation.
Thus, the electronic state in the domain has localized nature.

However, the electronic states in the domain are different from those in the Mott state in TaS$_2$ with respect to several points.
One is the finite DOS at Fermi energy ($E_{\mathrm{F}}$). The tunneling spectrum shown in Fig. 3(c) indicates the finite DOS at $E_{\mathrm{F}}$ whereas no DOS at $E_{\mathrm{F}}$ exists in the Mott state in TaS$_2$ \cite{}.
Another is the reduction of the Mott gap.
Although the value of the Mott gap is approximately 400 meV, the observed gap is approximately 200 meV.
One possible explanation for these differences is as a concequence of disorder. 
Theoretical calculation indicates that structural disorder induces the finite DOS at $E_{\mathrm{F}}$ and reduction of the effective Mott gap\cite{CuIntercalation}.
Another possible explanation is the decrease in the coherence of CDW order due to the existence of the walls, which we discuss later.

Figure 3(c) shows the tunneling spectrum obtained at the wall.
In contrast to the spectrum at the domain, the tunneling spectrum obtained at the wall does not show the peak at -100 mV.
Because the peak is reminiscent of the Mott gap, the lack of the peak at -100 mV indicates the suppression of the Mott state.
Although a peak structure at approximately 100 mV remains, the width of the peak becomes broader than that of the domain 
and the corresponding energy is higher than that of the domain.
These spectral differences suggest that the peak structure has some kind of different origin from the remnants of the Mott gap observed in the domain
but is due to a resonant state or a band across $E_{\mathrm{F}}$. 
Figure 3(e) shows the change in the tunneling spectra along the domain wall shown in Fig. 3(b).
In contrast to the domain, as can be seen in Fig. 3(e),
the conductance at near +100 mV does not show a periodic modulation, indicating the uniform electronic density along the wall.


The observed domain structure in Fe doped TaS$_2$ has several similarities to the ``metallic mosaic" produced by the application of bias voltage pulse \cite{pulseSTM1, pulseSTM2}.
The size of each domain and tunneling spectra at each region resemble each other.
Thus, the appearance of the domain structure is a universal feature at the melting of the Mott state in 1$T$-TaS$_2$.
However, there are several differences between them, specifically the stability of the domain structure.
The ``metallic mosaic" was reported to disappear with time and as temperature increases to 46 K.
However, we observed the same domain structure for at least a few days.
Moreover, STM observation on a sample with $x$ = 0.01 at 77 K also revealed the existence of the domain structure with size of each domain similar to that at 4.2 K.
These facts indicate that the domain structure in Fe doped TaS$_2$ is more stable than the ``metallic mosaic."

A previous STM study of Se substituted TaS$_2$ revealed the existence of a similar domain structure, and suggested that the domain structure results from stacking of distorted triangular lattices of the David-stars\cite{TaSSe_STM}.
In Se substituted TaS$_2$, the domain structure was observed at approximately 15 \% substitution,
whereas in Fe doped TaS$_2$, only 1 \% Fe substitution induces the domain structure.
Thus, the stacking of the disordered lattice is not likely to the origin of the domain structure in Fe doped TaS$_2$, but a rather local effect may cause the domain structure in Fe doped TaS$_2$.
Because ionic radius of Fe is smaller than that of Ta, structural distortion caused by the difference in the ionic radius may induce the domain structure near the Fe site. 
Another role of Fe on the domain structure is pinning of the walls.
Once the domain wall was formed, Fe site possibly pins the wall, resulting in the stabilization of the domain structure up to 77 K.
The difference in the valence number of Ta and Fe also seems to play a role in the formation of the domain structure.
However, we could not find any spectroscopic signature that derived from the difference at present.

\begin{figure}[tb]
\begin{center}
\includegraphics[width=8cm]{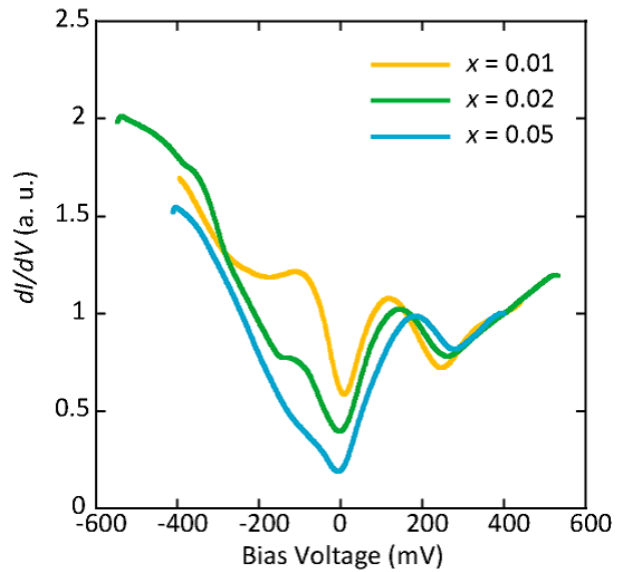}
\end{center}
\caption{
(Color online) Spatially averaged tunneling spectra of the sample with $x$ = 0.01, 0.02, and 0.05.
Each spectrum is normalized at 400 mV.
As increasing Fe concentration, $dI/dV$ at -100 mV reduces and the energy of the peak structure at positive bias increases.
}
\label{fig4}
\end{figure}

Finally, we make a comment on the emergence of superconductivity in this system.
Figure 4 shows spatially averaged tunneling spectra of the samples with $x$ = 0.01, 0.02, and 0.05.
With increasing Fe concentration, $dI/dV$ conductance at approximately -100 mV decreases.
Because, as shown in Fig. 3(c), the tunneling spectrum at the wall has no peak at approximately -100 mV, 
the gradual decrease in $dI/dV$ at -100 mV indicates the increase in the number of the David-stars that have a electronic state similar to that at the domain wall.
In fact, many fragmented domain walls can be seen in the sample with $x$ = 0.02 as shown in Fig. 2(c).
Since the superconductivity emerges at $x$ = 0.02, the increase in the wall regions seems to closely relate to the appearance of the superconductivity.
The appearance of the domain wall also reported in some transition metal dichalcogenides, such as Pt doped IrTe$_2$ and Cu intercalated TiSe$_2$ \cite{IrPtTe2_STM, IrPtTe2_STM2, CuTiSe2_STM, CuTiSe2_XRD}.
In TiSe$_2$, the CDW is three-dimensional, which is different from that in TaS$_2$.
Furthermore, Cu is intercalated between Se layers, whereas Fe is substituted in each Ta layer in TaS$_2$.
Despite these differences, however, both materials show domain walls where the phase of the CCDW changes sharply,
and show superconductivity at the concentration where the domain walls appear.

There are some possible roles of the walls in the appearance of the superconductivity in Fe doped TaS$_2$.
One possibility is that the superconductivity appears at the walls,
because the electronic state in the wall has a different nature from that of the domain,
which shows the localized nature, as shown in Fig. 3(e).
The other is that the existence of the wall reduces the coherence of the CCDW.
As the CCDW order is strongly tied to the Mott state in 1$T$-TaS$_2$,
the wall at which the phase of the CCDW changes sharply reduces the coherence of the CDW and localized nature in the domain.
Such a spatial change caused by the wall seems to result in the appearance of superconductivity.
Unfortunately, because of the measuring temperature of 4.2 K, the superconducting gap was not measured in this experiment.
The measurements of the spatial distribution of the superconducting gap is a future work.

In summary, we performed STM/STS measurements on 1$T$-Ta$_{1-x}$Fe$_{x}$S$_2$ for $x$ = 0, 0.01, 0.02, and 0.05.
Systematic STM measurements at 4.2 K revealed the evolution of the CDW state with Fe substitution.
The regular array of the David-stars in the pristine sample changes into the domain structure with Fe doping.
The STS measurement with high spatial resolution revealed that the domain structure consists of domains with localized nature and domain walls.
The domain structure was close to the previously reported ``metallic mosaic," 
which can be created by an application of the pulse voltage from an STM tip, except for the stability against temperature.
Because Fe doping increased the domain walls, it is thought that the domain wall is responsible for the appearance of the superconductivity.
These findings provide significant insights into the understanding of the melting of the Mott state and self-organization of the electronic state in 1$T$-TaS$_2$.

\footnotesize{{\textbf{Acknowledgement}}
This work was supported by JSPS KAKENHI Grant Number 17J08573.}


\nocite{*}
\thebibliography{99}

\bibitem{Hasegawa} J.-J. Kim, W. Yamaguchi, T. Hasegawa, and K. Kitazawa, Phys. Rev. Lett. \textbf{73}, 2103 (1994).
\bibitem{NCCDW_STM} X. L. Wu and C. M. Lieber, Phys. Rev. Lett. \textbf{64}, 1150 (1990).
\bibitem{Pressure} B. Sipos, A. F. Kusmartseva, A. Akrap, H. Berger, L. Forr\'{o}, and E. Tuti\v{s}, Nat. Mater. \textbf{7}, 960 (2008).
\bibitem{FeDoping} L. J. Li, W. J. Lu, X. D. Zhu, L. S. Ling, Z. Qu, and Y. P. Sun, Euro. Phys. Lett. \textbf{97}, 67005 (2012).
\bibitem{ARPES} R. Ang, Y. Tanaka, E. Ieki, K. Nakayama, T. Sato, L. J. Li, W. J. Lu, Y. P. Sun, and T. Takahashi, Phys. Rev. Lett. \textbf{109}, 176403 (2012).
\bibitem{SeDoping} Y. Liu, R. Ang, W. J. Lu, W. H. Song, and Y. P. Sun, App. Phys. Lett. \textbf{102}, 192692 (2013).
\bibitem{laser1} L. Stojchevska, I. Vaskivskyi, T. Mertelj, P. Svetin, S. Brazovskii, and D. Mihailovic, Science \textbf{344}, 177 (2014).
\bibitem{laser2} I. Vaskivskyi, J. Gospodaric, S. Brazovskii, D. Svetin, P. Sutar, E. Goreshnik, I. A. Mihailovic, T. Mertelj, D. Mihailovic, Sci. Adv. 1:e1500168 (2015).
\bibitem{pulseSTM1} L. Ma, C. Ye, Y. Yu, X. F. Lu, X, Niu, S. Kim, D. Feng, D. Tom\^{a}nek, Y -W. Son, X. H. Chen, and Y. Zhang, Nat. commun. \textbf{7}, 10956 (2016).
\bibitem{pulseSTM2} D. Cho, S. Cheon, K -S. Kim, S -H. Lee, Y -H. Cho, S -W. Cheong, H. W. Yeom, Nat. Commun. \textbf{7}, 10453 (2016).
\bibitem{AndersonLocalization} F. J. Di Salvo, J. A. Wilson, and J. V. Waszczak, Phys, Rev. Lett. \textbf{36}, 885 (1976).
\bibitem{STMTaFeS2} H. Chen, X. L. Wu, and C. M. Lieber, J. Am. Chem. Soc. \textbf{112}, 3326 (1990).
\bibitem{CuIntercalation} E. Lahoud, O. N. Meetei, K. B. Chaska, A. Kanigel, and N. Trivedi, Phys. Rev. Lett. \textbf{112}, 206402 (2014).
\bibitem{TaSSe_STM} T. Endo, W. Yamaguchi, O. Shiino, T. Hasegawa, and K. Kitazawa, Sur. Sci. \textbf{453}, 1 (2000).
\bibitem{IrPtTe2_STM} Y. Fujisawa, T. Machida, K. Igarashi, A. Kaneko, T. Mochiku, S. Ooi, M. Tachiki, K. komori, K. Hirata, H. Sakata, J. Phys. Soc. Jpn. \textbf{84}, 043706 (2015).
\bibitem{IrPtTe2_STM2} Y. Fujisawa, T. Machida, K. Igarashi, A. Kaneko, T. Mochiku, S. Ooi, M. Tachiki, K. komori, K. Hirata, H. Sakata, Physica C \textbf{530}, 35 (2016).
\bibitem{CuTiSe2_XRD} A. Kogar, G. A. de la Pena, S. Lee, Y. Fang, S. X.-L. Sun, D. B. Lioi, G. Karapetrov, K. D. Finkelstein, J. P. C. Ruff, P. Abbamonte, S. Rosenkranz, Phys. Rev. Lett. \textbf{118}, 027002 (2017).
\bibitem{CuTiSe2_STM} S. Yan, D. Iaia, E. Morosan, E. Fradkin, P. Abbamonte, and V. Madhavan, Phys. Rev. Lett. \textbf{118}, 106405 (2017).
\bibitem{VESTA} K. Momma, F. Izumi, J. Appl. Crystallogr. \textbf{44}, 1272 (2011).
\end{document}